\documentclass[prc,aps,floatfix,groupedaddress,amsmath,amssymb]{revtex4}
\usepackage{graphicx}
\usepackage{epsfig}
\usepackage{dcolumn}

\def\beq{\begin{equation}}
\def\eeq{\end{equation}}

\newcommand{\kv}{\mathbf{k}}

\newcommand{\kd}{{\rm k}}

\newcommand{\be}{\begin{eqnarray}}
\newcommand{\ee}{\end{eqnarray}}

\newcommand{\een}{\nonumber\end{eqnarray}}

\begin{document}

\title{Elementary excitations in homogeneous superfluid neutron star matter: \\
role of the neutron-proton coupling.}

\author{Marcello Baldo$^1$ and Camille Ducoin$^2$}
\affiliation{
$^1$ Dipartimento di Fisica, Universit\`a di Catania,
and INFN, Sezione di Catania,
Via Sofia 64, I-95123, Catania, Italy \\
$^2$ IPNL, CNRS/IN2P3, Universit\'e de Lyon/UCBL, F69622 Villeurbanne Cedex, France}

\begin{abstract}
The thermal evolution of Neutron Stars is affected by  the elementary excitations that characterize the stellar matter. In
particular, the low-energy excitations, with a spectrum linear in momentum, can play a major
role in the emission and propagation of neutrinos. In this paper, we focus on the elementary modes in the region of proton superfluidity,
where the neutron component is expected to have a very small or zero pairing gap.
 We study the overall spectral functions of protons, neutrons and electrons on the basis of the Coulomb and nuclear interactions.
 This study is performed in the framework of the Random Phase Approximation, generalized in order to
describe the response of a superfluid system. The formalism we use ensures that the Generalized Ward's Identities
are satisfied. Despite their relative small fraction, the protons turn out to modify the neutron spectral function as a consequence of the nuclear neutron-proton interaction. This effect is particularly evident at the lower density, just below the crust for a density close to the saturation value, while at increasing density the neutrons and the protons are mainly decoupled. The proton spectral function is characterized by a pseudo-Goldstone mode below $ 2\Delta $, twice the pairing gap, and a pair-breaking mode above  $ 2\Delta $. The latter merges in the sound mode of the normal phase at higher momenta. The neutron spectral function develops a collective sound mode only at the higher density. The electrons have a strong screening effect on the proton-proton interaction at the lower momenta, and decouple from the protons at higher momenta.    
\end{abstract}

\maketitle

\section{Introduction}

In Neutron Stars the elementary excitations of the matter affect the whole thermodynamics and long term 
evolution of the star, because they are involved in the emission and scattering of neutrinos, in the values of the heat content and 
transport coefficient.  
At not too high density it is expected that the main components of the matter are neutron, protons, electrons 
and muons~\cite{shap}, and then
the spectral properties of these excitations can have a complex structure.
Collective modes in asymmetric nuclear matter have been studied previously,
e.g. in Refs.~\cite{Haensel-NPA301, Matera-PRC49, Greco-PRC67}.
In the astrophysical context, a study of the collective excitations in normal neutron star matter
on the basis of the relativistic mean field method has been presented in Ref.~\cite{Providencia-PRC74}.
The spectral functions of the different components in normal neutron star matter
have been calculated in Ref.~\cite{paper1,paper2} on the basis of non-relativistic Random Phase Approximation (RPA)
for the nucleonic components and relativistic RPA for the leptonic components.
Different models for the nuclear effective interaction were considered
and a detailed comparison was done between some Skyrme forces and a microscopically derived interaction.
This work was extended to superfluid proton matter in ref. \cite{paper3}, but neglecting the proton-neutron coupling.
The elementary excitations in superfluid neutron star matter have been studied by several authors
~\cite{Reddy,Kundu,Leinson1,Armen,Leinson2,Vosk}.
In ref. \cite{paper3} it was shown that the proton superfluid presents a pseudo-Goldstone mode at low momentum, which can have a strong influence e.g. on neutrino emission~\cite{Yako,Reddy,Kundu,Leinson1,Armen,Leinson2,Vosk} or mean free path.
More recently in refs. \cite{Urban1,Thesis} the elementary excitations of superfluid neutron matter in the crust region were studied, and with the possible inclusion of the coupling with the nuclear lattice in refs. \cite{Chamel,Urban2,Thesis,Koby} .    
\par
In this paper we focus on the region of homogeneous matter where proton superfluidity is expected to occur, while the neutron component is predicted to have very small or zero pairing gap. This restricts the density region between about saturation density and twice saturation density. This work extends the study of refs. \cite{paper3} by including the neutron-proton interaction. An extensive study of the elementary excitations in presence of both proton and neutron superfluidity has been presented in ref. \cite{Koby1}, where the hydrodynamics formalism was used with the inclusion of proton-neutron coupling. 
As in ref. \cite{paper3} we formulate the general theoretical scheme within the generalized RPA approximation, which is known to be a conserving approximations~\cite{BaymKad,Baym}, i.e. current is conserved locally and the related Generalized Ward's Identities (GWI)~\cite{Schriefferb} are fulfilled. One of the main goal of the work is the study of the mutual influence of proton and neutrons on the overall spectral functions.
\par
The plan of the paper is as follows. In Sec. \ref{sec:form} the formalism for the response function in the generalized RPA scheme is briefly sketched. In particular it is discussed the method to estimate microscopically the effective nucleon-nucleon interaction. In Sec. \ref{sec:res} the results are presented for the spectral function taking the proton pairing gap as a parameter. The role of the neutron-proton interaction is discussed in detail. In Sec. \ref{sec:conc} we summarize and draw the conclusions. Finally in the Appendix additional details of the calculations are given. 
\section{Formalism. \label{sec:form}}
For future reference we sketch in this section the formalism of the generalized Random Phase Approximation (RPA). In a multi-components fermion systems the equations for the generalized response functions $ \Pi $ can be written schematically \cite{paper3,Schriefferb}
\beq
\Pi_{ik}(t,t') \,=\, \Pi_{ik}^{0}(t,t') \,+\, \sum_{jl} \,  \Pi_{ij}^{0}(t,\overline{t_1}) v_{j,l} \Pi_{lk}(\overline{t_1},t') 
\label{eq:RPA}\eeq
\par\noindent
where $ \, i,j, \cdots \,$ label the different components and the corresponding degrees of freedom, $\,  v_{j,l}\, $ is the effective interaction between them and $\, \Pi^{0}\, $ is the free response function. The time variable with an overline is integrated.
Specifically, for the considered Neutron Star matter one has neutron, proton and electron components (neglecting muons),
with the possibility of both particle-hole and pair excitations in the proton channel. If we take into account both proton and neutron pairing, in terms of creation and annihilation operators the indexes $\, i,j, \cdots $ run over the following configurations
\beq
\begin{array}{ll}
\ &\ a^\dag(p)\,a(p) | \Psi_0>\,\,\ ,\ \,\, a^\dag(p)\,a^\dag(p) | \Psi_0>\,\,\ ,\ \,\, a(p)\,a(p) | \Psi_0> \\
\ &\ \\
\ &\ a^\dag(n)\,a(n) | \Psi_0>\,\,\ ,\ \,\, a^\dag(n)\,a^\dag(n) | \Psi_0>\,\,\ ,\ \,\, a(n)\,a(n) | \Psi_0> \\ 
\ &\ \\
\ &\ a^\dag(e)\,a(e) | \Psi_0>  
\end{array}
\label{eq:conf}\eeq
\par\noindent
where the labels $\, n, p, e\, $ indicate neutrons, protons and electron respectively, and $ |\Psi_0> $ is the correlated ground state. If we call $ A_i $ the generic configuration, the response functions can be written
\beq
 \Pi_{ik}(t,t') \,=\, - < \Psi_0 | T\{A_i^\dag(t) A_k^{\phantom{\dag}}(t') | \Psi_0 > 
\label{eq:Pi}\eeq 
\par\noindent
where $ T $ is the usual fermion time ordering operator.
The configurations (\ref{eq:conf}) correspond in fact to both density and pairing excitations. In agreement with (\ref{eq:conf}), in principle Eqs. (\ref{eq:RPA}) form a $ 7\times 7 $ system of coupled equations. However, as mentioned in the Introduction, neutrons are assumed to be in the normal state. Furthermore, it turns out \cite{paper3,Schrieffer} that one equation can be decoupled to a good approximation by taking suitable linear combinations of the pairing additional mode $ a^\dag(p)\, a^\dag(p) $ and pairing removal mode $ a(p)\, a(p) $. In this way the system reduces to  $ 4\times 4 \, $ coupled equations. Details on the equations and their explicit analytic form are given in Appendix A. The system has to be solved for the response functions $\, \Pi_{ik}\, $, all of which can be obtained by selecting the inhomogeneous term in Eq. (\ref{eq:RPA}). More precisely one has to select a given configuration indicated by the right index $ k $ and solve the system for each choice of $ k $. In this way one gets all the diagonal and non-diagonal elements of $\, \Pi_{ik}\, $.  
\par
One has to notice that the generalized RPA equations are valid in the collisionless regime, so that the only damping of the modes is the Landau damping, which is very effective above a certain momentum threshold. In particular dissipation due to electron-electron collisions is neglected. This is justified if the electron mean free path is much larger than the typical wavelength of the mode. Under the physical conditions of Neutron Star matter, i.e. low temperature and density of the order of the saturation one, the electron mean free path was estimated in ref. \cite{Shtern}, where it was shown that the collisions are dominated by the exchange of transverse plasmon modes and the mean free path extends to a macroscopic size, of the order of 10$^{-3}$ cm.
This also indicates that the collision time is much longer than the characteristic period of the modes, and that the electron collisions are relevant only for macroscopic motion like viscous flow. We therefore neglect in the sequel electron dissipation, and include only the Landau damping.           
\par 
For simplicity for the single particle energy spectrum we are going to use the free one. The introduction of the effective masses is trivial and we think that it is not going to change qualitatively the overall pattern of the results, but of course for quantitative results the effective mass is mandatory. Then the main input needed in Eq. (\ref{eq:RPA}) is the effective interaction $ v_{ij} $. The proton pairing interaction strength $ U $ is very sensitive to many-body effects \cite{ppair} and it is quite challenging to estimate its size. We prefer to use the proton pairing gap as a parameter to be explored and fix the pairing interaction consistently with the gap equation  ( $ U > 0 $ )
\be
\Delta &=& U\int{\frac{d^3 \kv}{(2\pi)^3}\frac{\Delta}{2E_{\kd}}}
= U\int{\frac{d^3 \kv}{(2\pi)^3} u_{\kd}v_{\kd}}
\label{eq:gap}\ee
\noindent The quasi-particle energy $ E $  and coherence factors  $ u, v $ have the standard form
\be
\begin{array}{ll}
 E_{\kd} &\,=\, \sqrt{(\epsilon_{\kd} - \mu)^2 + \Delta^2} \\
\ &\ \\
 v_{\kd}^2 &\,=\, \frac{1}{2} \big( 1 \,-\, \frac{\epsilon_{\kd} - \mu}{E_{\kd}} \big) 
 \ \ \ \ \ , \ \ \ \ \   u_{\kd}^2 \,=\, \frac{1}{2} \big( 1 \,+\, \frac{\epsilon_{\kd} - \mu}{E_{\kd}} \big) \\  
\end{array}
\label{eq:stand}\ee
\noindent where $ \epsilon_{\kd} $ is the proton kinetic energy and $ \mu $ the proton chemical potential. In the calculations the value of the pairing gap $ \Delta $ is fixed at a given value and the effective interaction strength $ U $ is extracted from the gap equation (\ref{eq:gap}). Then the pairing interaction $ -U $ is inserted in the RPA equations (\ref{eq:RPA}). \par 
For the various particle-hole interactions we focus on the density response function, i.e. the vector channel, and then we follow the Landau monopolar approximation. The corresponding strength can be estimated on the basis of a realistic Skyrme interaction.
It is also possible to consider microscopic many-body Equation of State (EOS) as an Energy Density Functional. In this approach for  Brueckner-Hartree-Fock (BHF) calculations the interaction strength can be obtained from the derivative with respect to the density $ \rho_j$ of the BHF potential energy $ V_i $, with $ i, j $ running on the proton and neutron components
\beq
\label{eq:vres-derU}
v_{ij}=\left(\frac{\delta V_i}{\delta\rho_j}(k_{{\rm F}i},\rho_n,\rho_p)\right)_{k,\rho_i={\rm cst}}\;.
\eeq
\noindent Notice that in Eq. (\ref{eq:vres-derU}) the Fermi momenta $ k_{{\rm F}i} $ must be kept fixed in performing the derivative.
How to do that in practice from the numerical calculations is explained in ref. \cite{paper1}. In the case of Skyrme forces the possible dependence on momenta is analytic and the procedure is trivial. In any case no strong pairing effects are considered,
e.g. the effective interactions are assumed to be independent of the pairing gap and calculated for the normal system.
\section{Results.\label{sec:res}}
\label{sec:results}
Before presenting the results obtained when all interactions are fully taken into account (namely nuclear pairing, Coulomb, and density-density nuclear interactions), it is instructive to discuss briefly the development of the overall structure of the excitation spectrum as the interactions are introduced. This can help to characterize the effects of each one of them.
\par
It is well known that in a neutral superfluid, with only pairing interaction, there are two types of excitations. Below $ 2\Delta $
a sharp Goldstone mode is present. It is a consequence of the breaking of gauge invariance that occurs in the ground state of a superfluid system, and its energy is linear in momentum for small momenta, with a velocity equal to $ v_F/\sqrt{3} $, being $ v_F $ the Fermi velocity. Notice that a gauge transformation on the field operators is equivalent to a $ U(1) $ transformation on the order parameter \cite{Greiter}.
At increasing momentum the energy spectrum deviates from linearity and approaches $ 2\Delta $ for large momenta \cite{paper2}. Above $ 2\Delta $      
another excitation mode appears, usually indicated as " pair breaking "  mode, because it corresponds indeed to the breaking of a Cooper pair. It is strongly damped and it is reflected in a bump of the spectral function above $ 2\Delta $. The energy of this mode, after increasing with the momentum, bends down towards $ 2\Delta $ also. At even higher momentum the spectral function has no structure and any excitation is overdamped \cite{paper2}. \par 
If we consider only the proton component and introduce the Coulomb interaction, in principle the Goldstone mode should disappear,
and it should be substituted by a proton plasma mode which has a finite energy at zero momentum. However it has been shown in ref. \cite{paper2} that the electrons are fully screening the proton-proton Coulomb interaction, and a sound mode reappears below $ 2\Delta $. The (screened) Coulomb interaction however affects the sound velocity, which turns out to be about three times the Goldstone mode velocity. This mode can be considered a pseudo-Goldstone mode, since it is still below $ 2\Delta $ but with a modified velocity due to the interaction. Details can be found in ref. \cite{paper2}, where the structure of the proton spectral function is discussed in detail. Here we notice only that the electron Thomas-Fermi screening length $ v_F / \sqrt{3}\omega_p $, where $\omega_p$ is the electron plasma frequency and $v_F$ the Fermi velocity, at saturation density and with the calculated proton fraction is about 18 fm and about 11 fm at twice saturation density. This is more than one order of magnitude larger than the average interparticle distance. However the Coulomb interaction is in any case screened, and this is enough to suppress the proton plasma mode. \par 
We now introduce the nuclear interaction, including the proton-neutron nuclear coupling. The proton fraction is taken from BHF calculations, which include three-body forces and correctly reproduce the phenomenological saturation point. The corresponding nuclear interaction strengths are calculated according to Eq. (\ref{eq:vres-derU}). The values of these physical parameters are reported in ref. \cite{paper1}. The comparison with Skyrme forces will be discussed later.\par 
The spectral or strength function $ S(q,\omega) $, at given momentum $ q $ and energy $ \omega $, is directly related to the response function $ \Pi $
\beq
S_i(q,\omega) \,=\, -{\rm Im} \big( \Pi_{ii}(q,\omega) \big) 
\label{eq:strength}\eeq
\noindent where $ {\rm Im} $ indicates the imaginary part and the index $ i $ runs over the particle-hole configurations of neutrons, protons and electrons. In Fig. 1a are reported the strength functions of the three components at the matter density $ \rho \,=\,$ 0.16 fm$^{-3}$, for which the proton fraction is 3.7\%, and for the momentum $ q \,=\, $0.025 fm$^{-1}$. The pairing gap is assumed to be 1 MeV \cite{ppair}. 
\begin{figure}
\vskip -8 cm
\includegraphics[bb= 300 0 470 790,angle=0,scale=0.55]{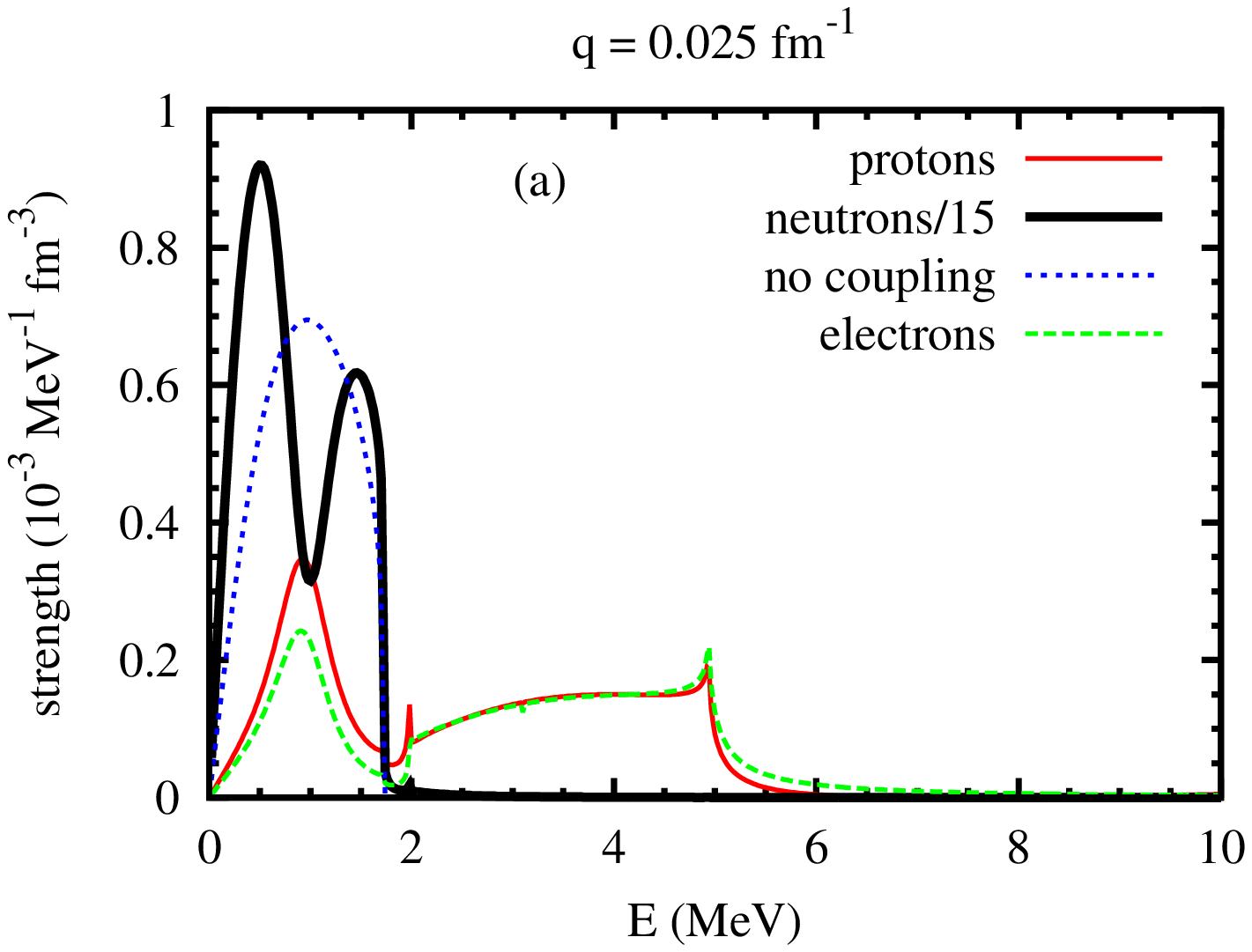}
\includegraphics[bb= 80 0 230 790,angle=0,scale=0.55]{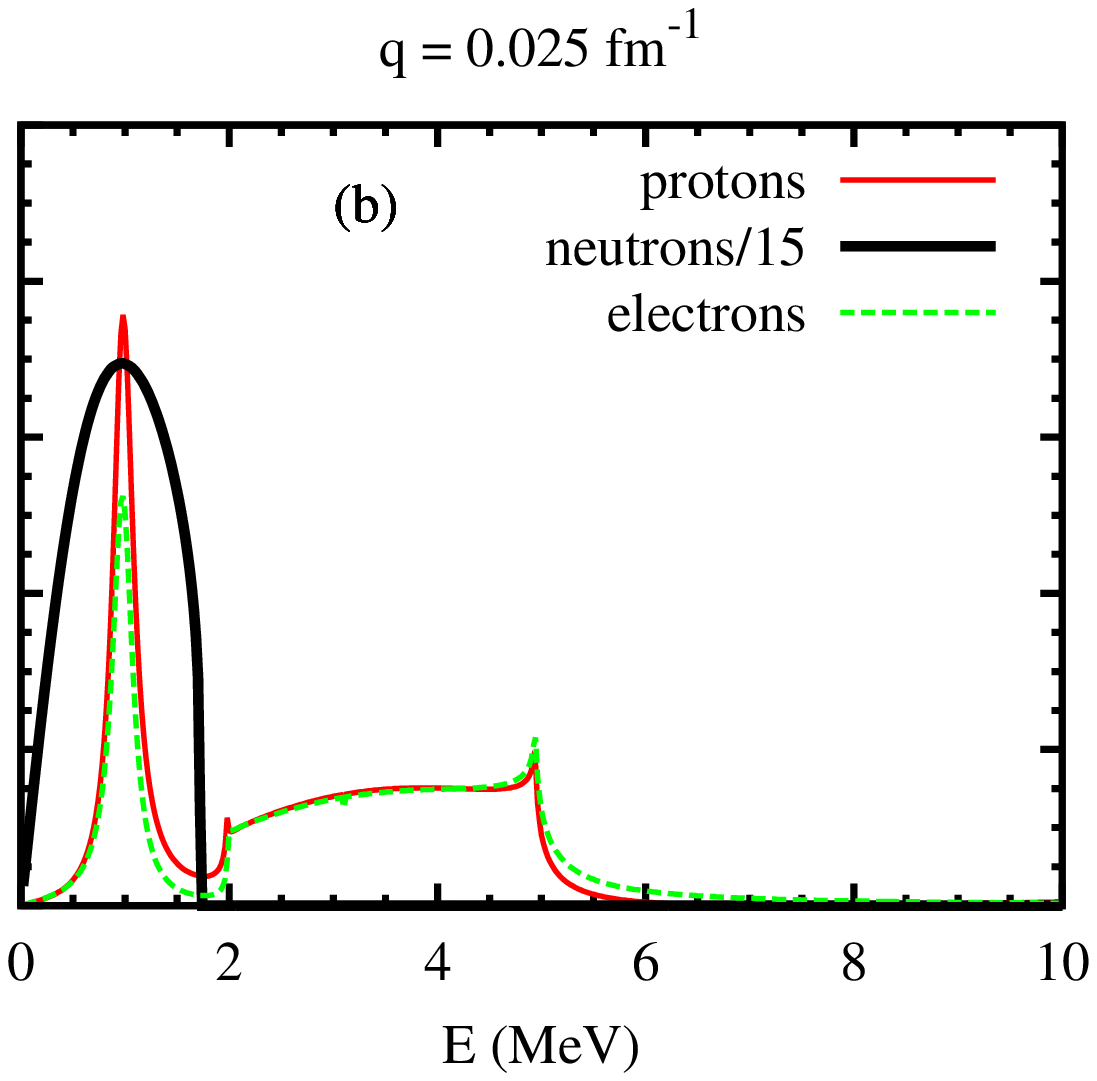}
\vskip -1 cm
\caption{(Color on line) Spectral functions of neutrons (black thick line), proton (red thin line) and electrons (green dashed line). 
In panels a) are reported the spectral functions when the Coulomb and nuclear interactions are introduced. 
For comparison in panel b) are reported the spectral functions when the neutron-proton interaction is neglected. For a closer comparison the neutron spectral function of this case is also reported in panel a) ( blue dotted line ) }
\label{fig:Fig1}
\end{figure}      

The neutron strength function (SF) is of course much larger than the proton one, due to the higher density of states, and for convenience it has been divided by 15. For comparison in Fig. 1b are reported the SF when the neutron-proton coupling is switched off. A few comments are in order. The proton SF (red thin line) is characterized by the pseudo-Goldstone mode below 2$\Delta$ and a pair-breaking mode above 2$\Delta$. Both modes persist when the neutron-proton interaction is switched on. However the presence of the neutron component suppresses the strength of the pseudo-Goldstone by about a factor two.
Furthermore  the pseudo-Goldstone  acquires an additional width, which is the consequence of the large spread of the neutron SF (black thick line). The neutron-proton coupling is apparent from the dip present in the neutron SF around the energy of the proton pseudo-Goldstone mode, and from the shift to lower energy of the neutron SF. Notice that the neutron SF, without proton coupling,  is smooth and unstructured. The reason is that the neutron effective particle-hole interaction is attractive in this density range, so that the possible sound mode lies inside the particle-hole continuum and it is therefore overdamped (Landau damping). In order to facilitate the comparison this uncoupled SF is reported also in Fig. 1a (blue dotted line).
Finally the electron SF (green dashed line) is following closely the proton one, which indicates the full electron screening of the proton-proton Coulomb interaction. \par 
At higher momenta the neutron-proton coupling is stronger, as it is apparent in Figs. 2a,b, where the spectral functions are displayed at the two higher momenta $ q \,=\, 0.05,\, 0.075 $\, fm$^{-1}$. In fact the neutron spectral function is strongly modified by the coupling with the proton component. Not only a dip is present around the energy of the proton pseudo-Goldstone, but the whole structure of the spectral function is strongly modified, which is a little surprising in view of the smallness of the proton fraction. It is also important to notice that the electron spectral function does not follow any more closely the proton spectral function, since at higher momenta the proton velocity starts to be comparable with the electron Fermi velocity.   
\begin{figure}
\vskip -8 cm
\includegraphics[bb= 300 0 470 790,angle=0,scale=0.55]{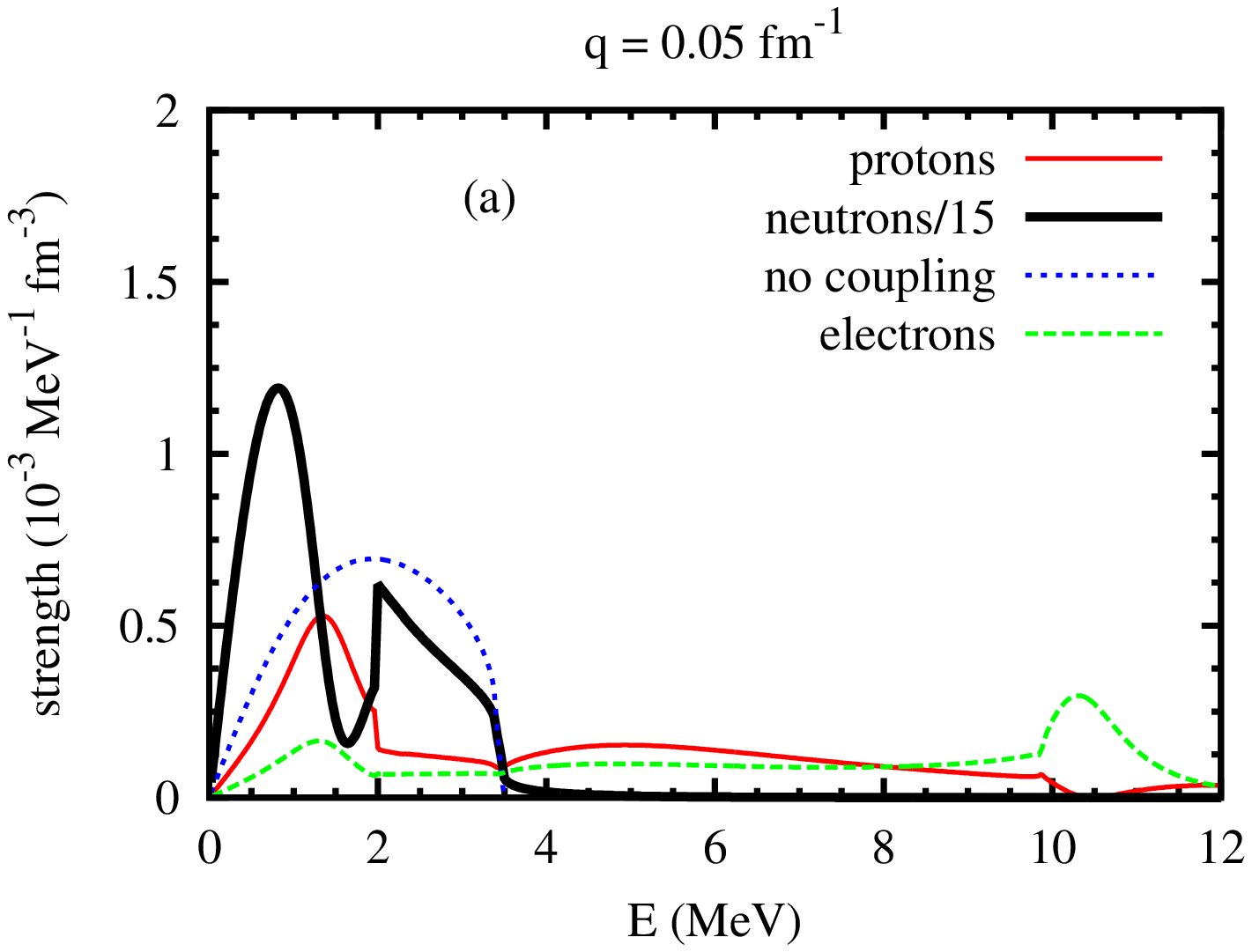}
\includegraphics[bb= 80 0 230 790,angle=0,scale=0.55]{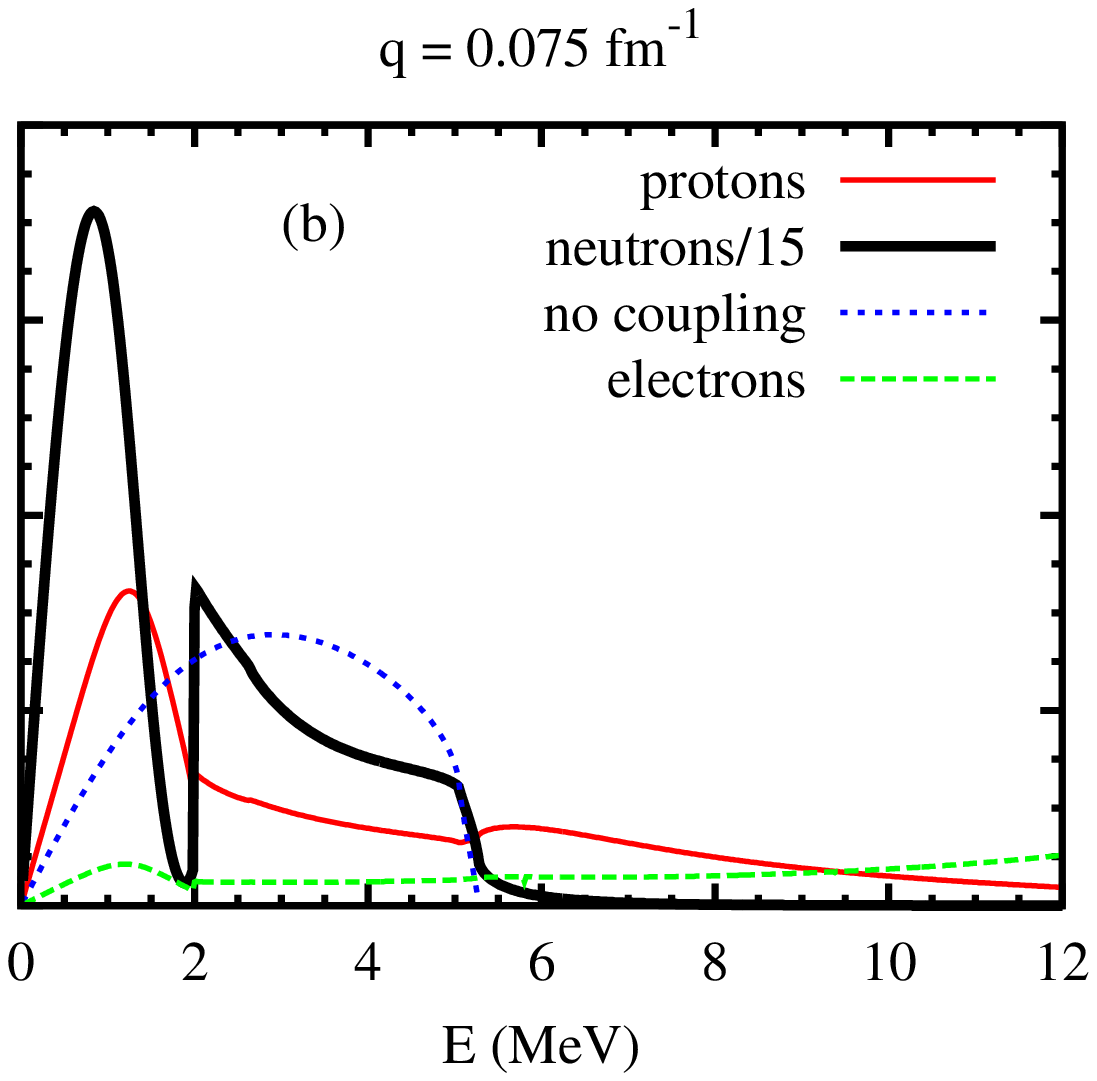}
\vskip -1 cm
\caption{(Color on line) Spectral functions as in Fig. 1a, for other two different momenta.}
\label{fig:Fig2}
\end{figure}
\par
It is interesting to compare the spectral functions in the case of proton superfluidity with the ones in the normal system.
We then fix the pairing gap at $ \Delta \,=\, $0.02 MeV. For not too small momenta the system is expected to behave as a normal system. In Figs. 3a,b are reported the spectral functions for two values of the momentum. If we compare Fig. 3a with Fig. 1a one notices a few apparent differences. First of all for the small value of the gap there is a sharp peak in the electron spectral function, corresponding to the electron plasmon excitation. This feature indicates that, as already noticed in ref. \cite{paper2}, the electron plasma mode is damped by the presence of a pairing gap. The reason of the damping is due to the opening of a new plasmon decay channel due to the coupling with the pair breaking mode through the Coulomb interaction. The peak in the proton spectral function looks similar. However the nature of the corresponding excitation is different. For this small value of the pairing gap the excitation energy is well above $ 2\Delta $ and the mode is a sound mode instead of a pseudo-Goldstone mode. However the neutron-proton coupling looks quite strong also in this case. Of course in this case no pair-breaking mode is present, as one can see from the absence of any appreciable proton strength at higher energy, in comparison with the large strength that appears just above $ 2\Delta $ in the case of $ \Delta \,=\, 1 $ MeV, as depicted in Fig. 1a. Notice that the electron plasmon peak does not appear in Fig. 3b because, at the indicated momentum, it lies outside the considered energy range. In any case at high enough momentum the electron plasmon is washed out by the Landau damping.         
\begin{figure}
\vskip -8 cm
\includegraphics[bb= 300 0 470 790,angle=0,scale=0.55]{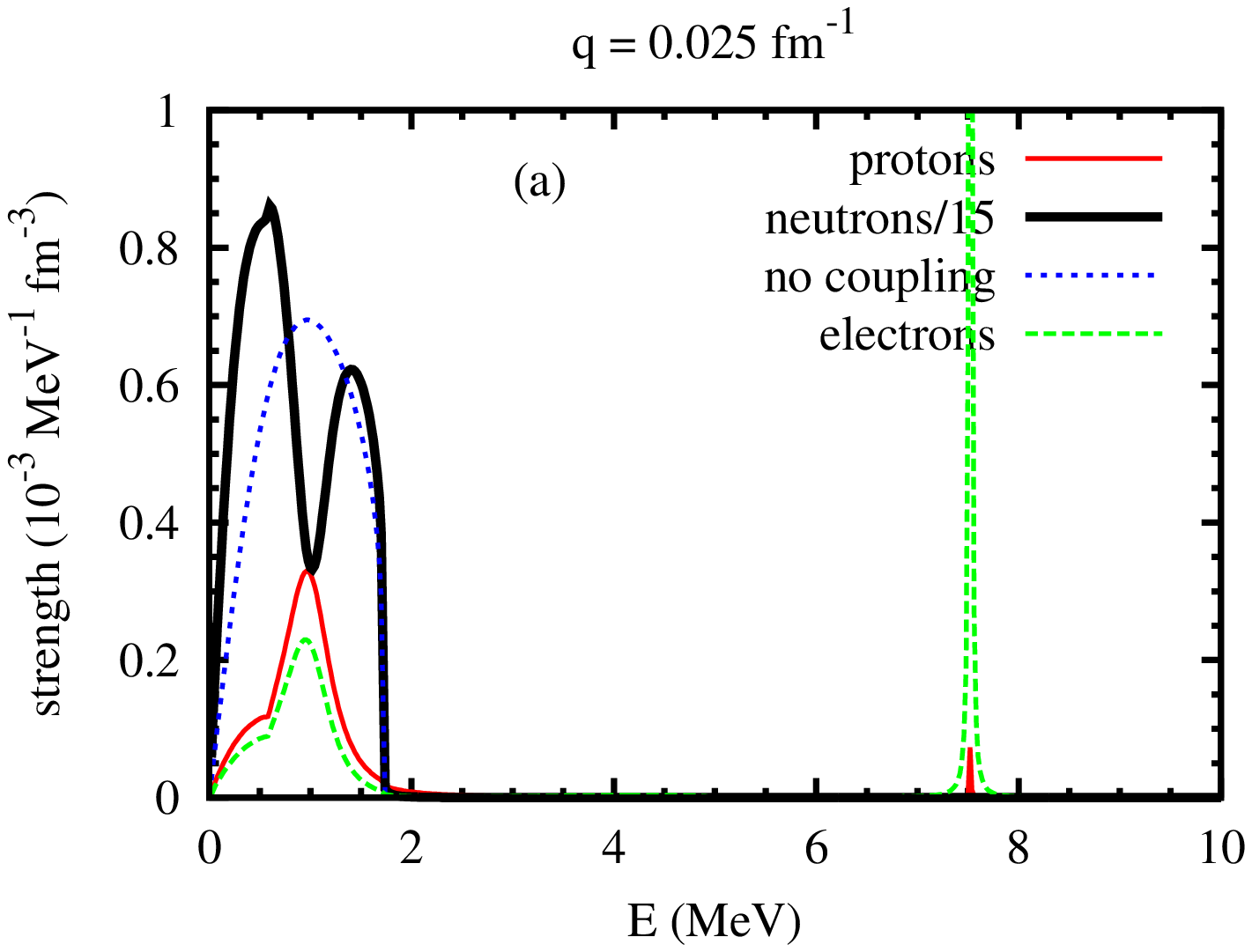}
\includegraphics[bb= 80 0 230 790,angle=0,scale=0.55]{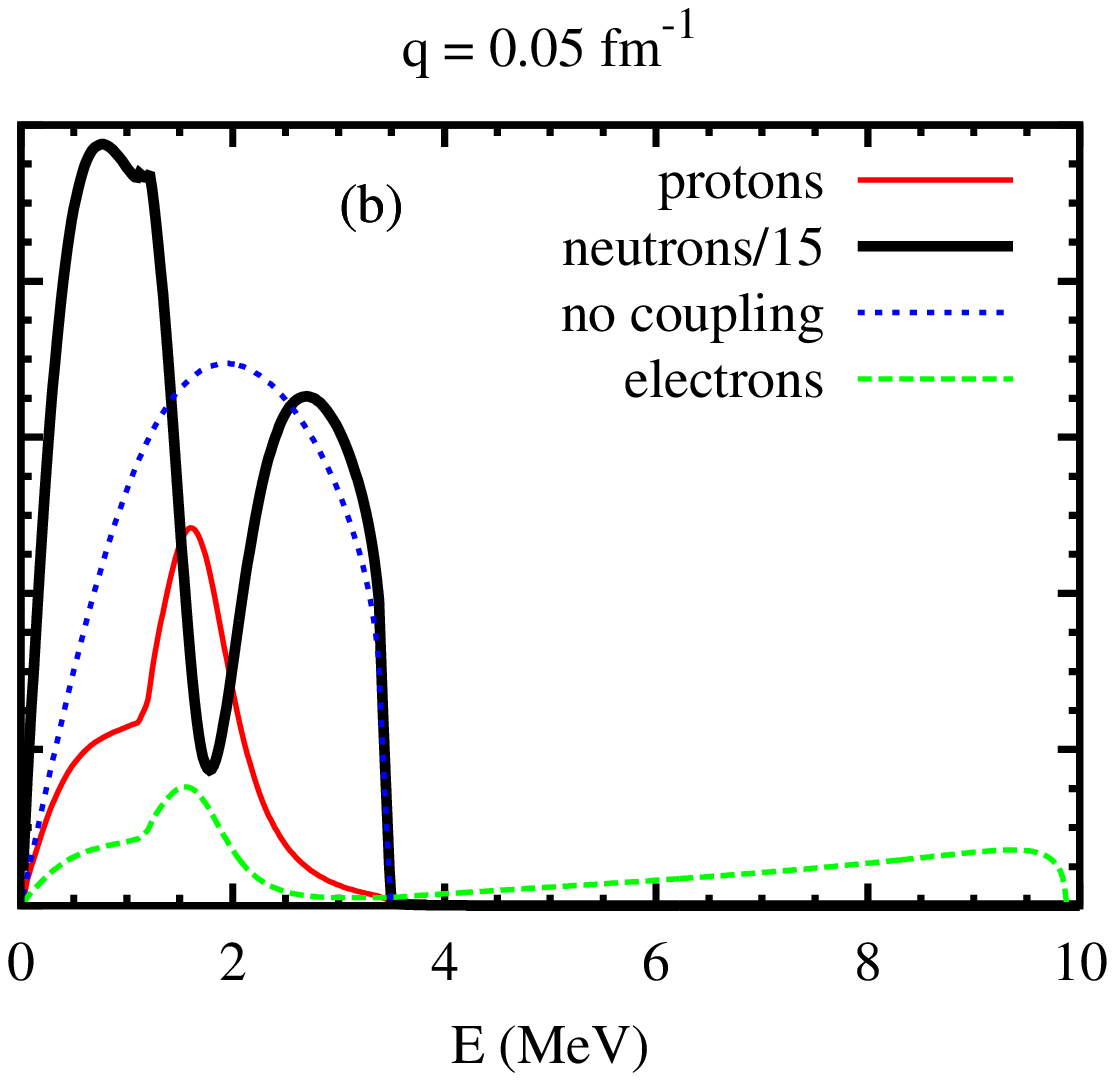}
\vskip -1 cm
\caption{(Color on line) Spectral functions for the small pairing gap $ \Delta \,=\, $0.002 MeV. Symbols like in Figs. 1,2.}
\label{fig:Fig3}
\end{figure}
\par
\begin{center}    
\begin{figure}
\vskip -7 cm
\includegraphics[bb= 200 0 360 790,angle=0,scale=0.5]{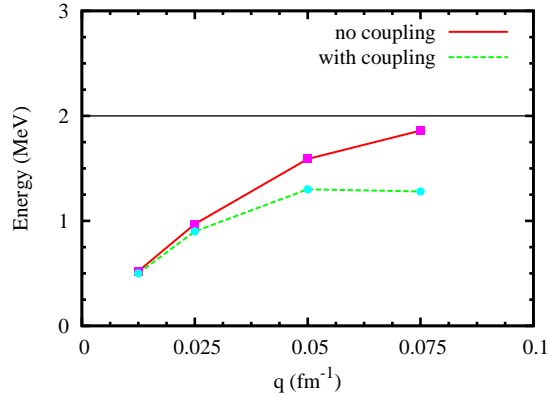}
\vskip -1. cm
\caption{(Color on line) Energy of the zero of the real part of the inverse of the response function at different momenta $ q $. The pairing gap is $ \Delta \,=\,$ 1 MeV. The red full line and full squares correspond to the case when the neutrons and the proton are uncoupled. The green dashed line and the full circles correspond to the case when the interaction between neutrons and protons is switched on.}
\label{fig:Fig4}
\end{figure}
\end{center}

It can be instructive to locate the position of the peak in the proton spectral function as a function of momentum. To be quantitative one can extract the position of the peak as the zero of the real part of inverse response function, that is the zero
of the real part of the matrix of Eq. (\ref{eq:RPA}). For a sharp mode this is indeed the position of the energy. Since the uncoupled neutron component does not support in this case any well defined mode, its effect is only to produce an additional width to the proton excitation mode. In Fig. \ref{fig:Fig4} is reported the energy of the peak so defined as a function of the momentum in the case of an uncoupled neutron-proton system and in the case of the coupled one. In the case of the uncoupled system one observes the expected \cite{paper2} smooth increase of the energy towards twice the pairing gap ($ \Delta \,=\,$ 1\, MeV).
At variance with this behavior, in the case of the coupled system the energy of the mode seems to saturate to a value which is a fraction of 2 $\Delta$. It has to be stressed however that the width of the excitation mode is quite substantial and the position of the peak is largely undefined. \par 
\par
\begin{figure}
\vskip -8 cm
\includegraphics[bb= 300 0 470 790,angle=0,scale=0.55]{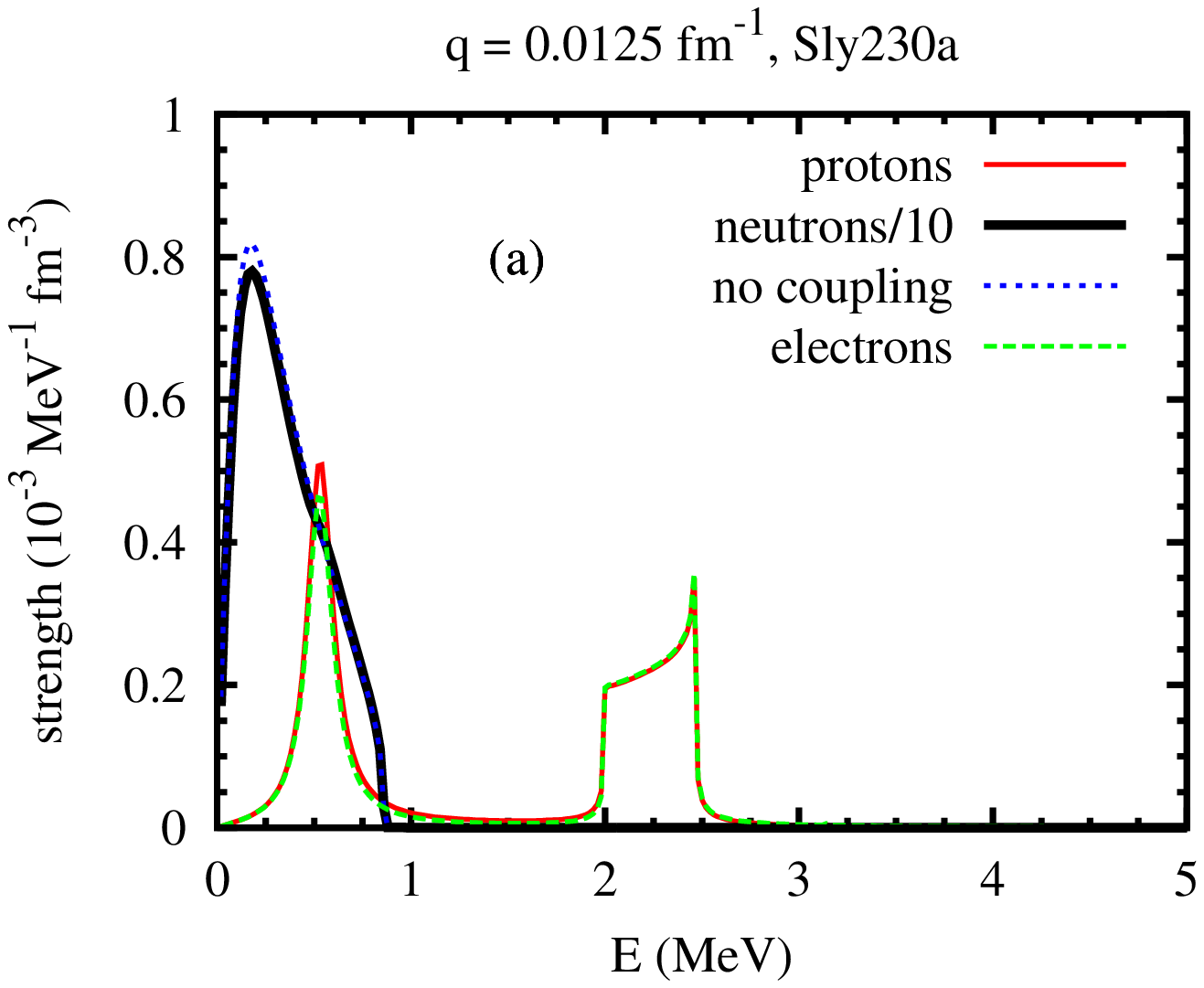}
\includegraphics[bb= 80 0 230 790,angle=0,scale=0.55]{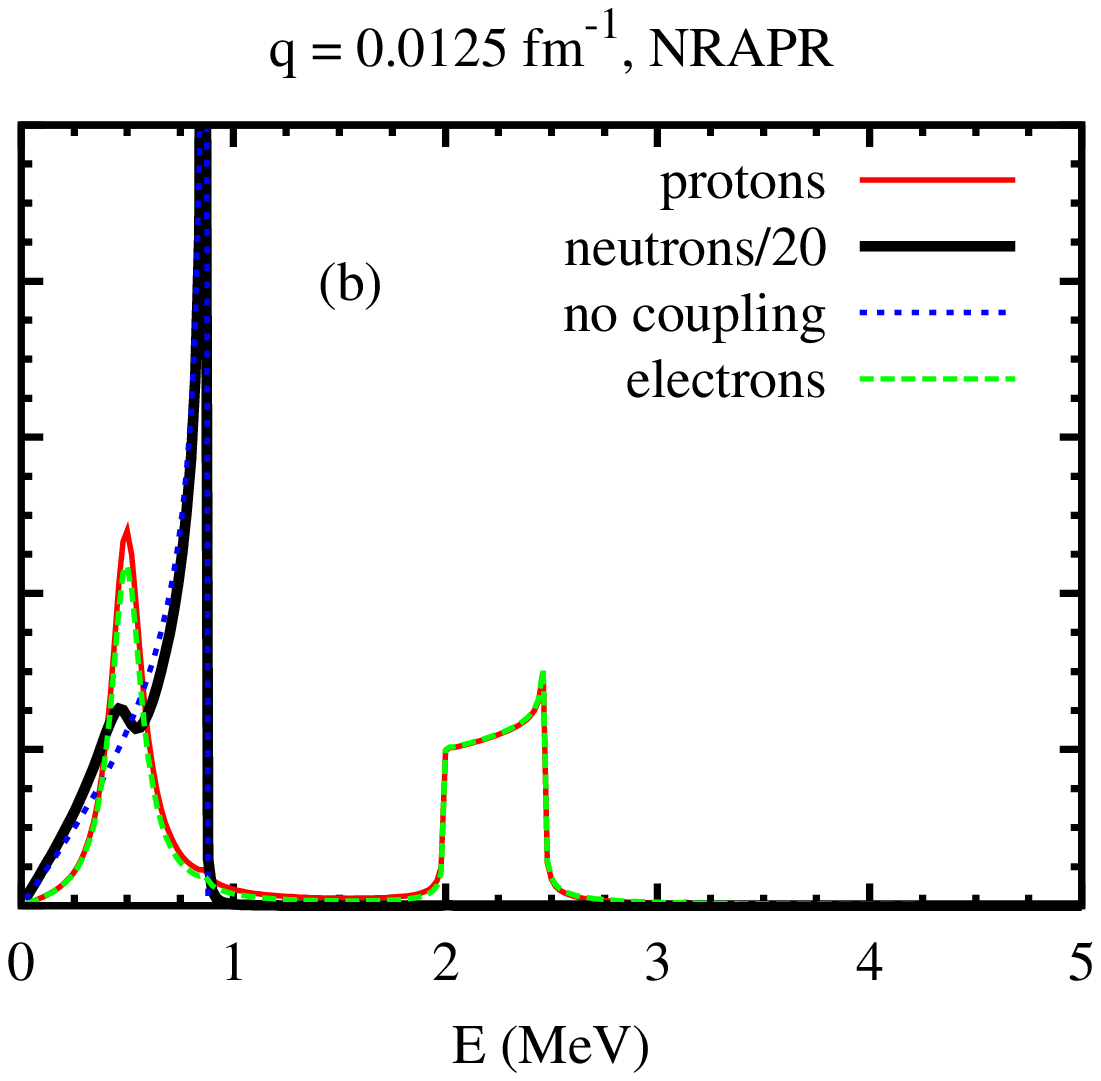}
\vskip -1 cm
\caption{(Color on line) Spectral functions for the Skyrme force Sly230, panel (a), and for the functional NRAPR, panel (b).  The pairing gap is $ \Delta \,=\, $1 MeV. Symbols as in Figs. 1,2,3.}
\label{fig:Fig5}
\end{figure}

It is interesting to compare these results, obtained within the BHF microscopic approach, with the ones that can be obtained with Skyrme functionals. In Fig. \ref{fig:Fig5} are reported the spectral functions for two Skyrme functionals, the Sly230a \cite{SLya} and the NRAPR \cite{NRAPR}. It is apparent that the neutron-proton coupling is substantially smaller in both cases with respect to the microscopic one. Notice anyhow that the Sly230a gives an attractive neutron-neutron interaction, while for the NRAPR it is repulsive.
In the latter case in fact a well defined neutron peak is present in the spectral function. This shows that the spectral function is quite sensitive to the choice of the Skyrme functional.\par
\begin{figure}
\vskip -8 cm
\includegraphics[bb= 300 0 470 790,angle=0,scale=0.55]{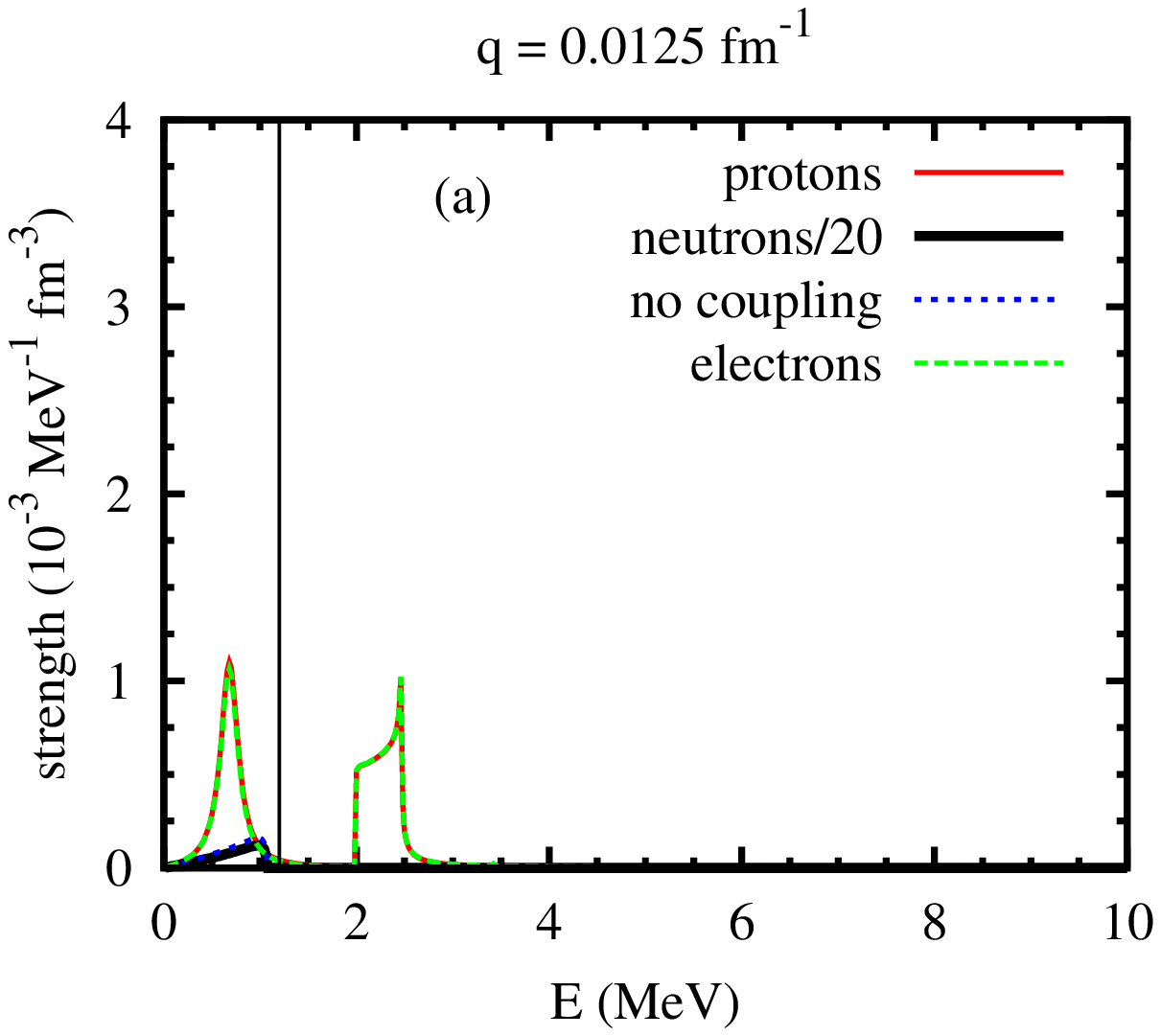}
\includegraphics[bb= 80 0 230 790,angle=0,scale=0.55]{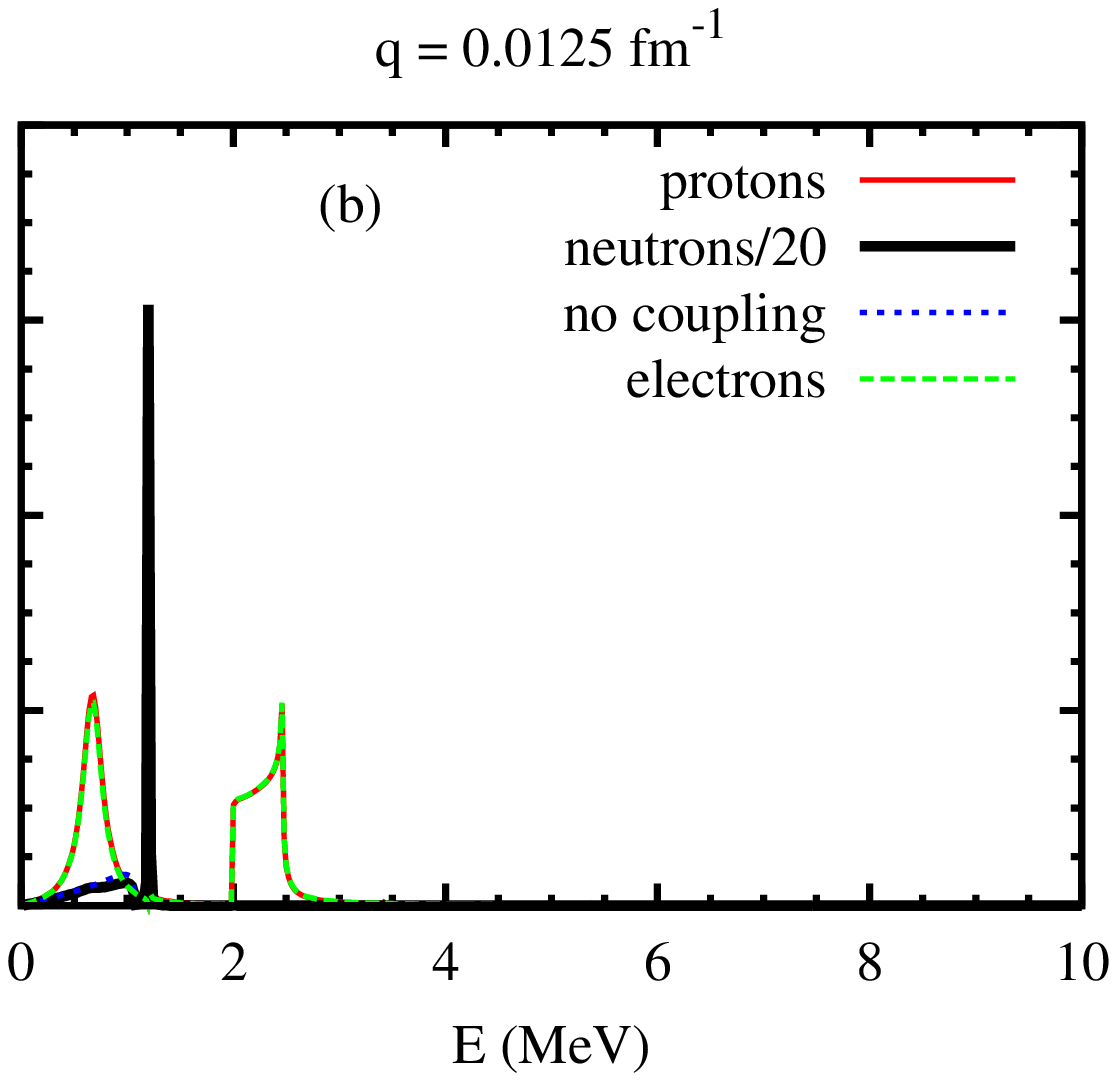}
\vskip -1 cm
\caption{(Color on line) Spectral function within the microscopic approach to the effective nuclear interaction without neutron-proton coupling at twice saturation density, panel (a). The same, but with the neutron-proton interaction included, panel (b).  The pairing gap is $ \Delta \,=\, $1 MeV. Symbols as in Figs. 1,2,3.}
\label{fig:Fig6}
\end{figure}

At twice saturation density the spectral function develops as illustrated in Fig. \ref{fig:Fig6} for the BHF estimate of the nucleon-nucleon interaction. In panel (a) is reported the spectral function for no neutron-proton coupling. At this density the neutron-neutron interaction is repulsive and a sharp delta-like peak is present for the neutron component. The picture changes only little when the neutron-proton coupling is switched on, panel (b). The only effect is the presence of a small width for the neutron peak,
but no neutron-proton coupling effect is apparent. This means that the proton and neutron component are excited independently, despite the fact that the proton fraction has increased to about $ 8\% $. 
It turns out also that the two Skyrme functionals give a repulsive neutron-neutron effective interaction at this density and the spectral functions are quite similar to the microscopic one.    

\par 
\section{Summary and conclusions. \label{sec:conc}}
In the region below the crust of a neutron star the dense matter is expected to be composed mainly by neutrons, protons and electrons. The collective elementary excitations of the medium are determined in general by the coupling among the three components and the corresponding spectral function has three components. We have studied the spectral function of the density excitations, separating the neutron, proton and electron strength functions. We focused on the density region where protons are expected to be superfluid, while neutrons are expected to be normal or with a very small pairing gap. The generalized RPA scheme in the Landau monopole approximation was used for the calculations of the density-density spectral function, which describes the response of each component to an external probe. A key quantity for this analysis is the nuclear interaction among the nucleons, which was derived by adopting the energy calculated in the Brueckner-Hartee-Fock  approximation as an energy density functional for nuclear matter. The microscopic  BHF calculation includes three-body forces and for symmetric matter it reproduces the phenomenological saturation point. Two Skyrme functionals were also considered for comparison. The inclusion of the electron-proton Coulomb interaction is essential, since the electron screening on the Coulomb proton-proton interaction  strongly affects the excitation branch corresponding to the pseudo-Goldstone mode associated with the proton superfluidity. Although the proton fraction is few percents, the neutron component of the spectral function is substantially affected by the neutron-proton interaction, while the proton component still presents a peak at the position of the pseudo-Goldstone mode. At twice saturation density one observes an essential decoupling between neutrons and protons,
both for the microscopic and the Skyrme functionals. In this case all the considered interactions give a repulsive neutron-neutron effective interaction and the neutron component of the spectral function displays a sharp peak corresponding to a well defined excitation mode.     \par
Finally, the calculations can be extended to the spin-density excitations, but in this case the modes should be much less collective,
as expected by the small value of the corresponding Landau parameter in symmetric matter \cite{Dick,BF}.   
\appendix
\section{}
\label{ap:detail}
\par
In this appendix we write down the explicit expressions of the system of equations (\ref{eq:RPA}) in the main text for the response functions $ \Pi_{ij} $. In extended form the system can be written as follows
\beq
\left(\begin{array}{cccc}
1-X^{pp}_{+}U_{\rm pair} & -2X_{GF}^{-}v_{pp} & 2X_{GF}^{-}v_c & -2X_{GF}^{-}v_{pn}  \\[0.25cm]
X_{GF}^{-}U_{\rm pair} & 1-2X^{ph}_{p}v_{pp}  & 2X^{ph}_{p}v_c  & -2X^{ph}_{p}v_{pn}\\[0.25cm]
      0            &  2X^{e}v_c & 1-2X^{e}v_c &  0  \\[0.25cm]
      0            &  2X^{ph}_{n}v_{np} &     0      & 1-2X^{ph}_{n}v_{nn} \\[0.25cm]        
\end{array}\right) 
\left(\begin{array}{c}
\Pi^{ (+)}_S\phantom{ph} \\[0.25cm]
\Pi^{(ph,p)}_S \\[0.25cm]
\Pi^{(ph,e)}_S\\[0.25cm]
\Pi^{(ph,n)}_S\\[0.25cm]
\end{array}\right)
= \left(\begin{array}{c}
\Pi^{(+)}_{0,S}\phantom{ph} \\[0.25cm]
\Pi^{(ph,p)}_{0,S} \\[0.25cm]
\Pi^{(ph,e)}_{0,S} \\[0.25cm]
\Pi^{(ph,n)}_{0,S} \\[0.25cm]
\end{array}\right)
\label{eq:RPA2}
\eeq
\noindent
Here we have introduced the notation:
\be
\label{eq:Xpp}
X_{+}^{pp}&=&\frac{1}{2}\left[X_{GG}^{pp}(q)+X_{GG}^{pp}(-q)\right] + X_{FF}(q) \\
\label{eq:Xph}
X_{p}^{ph}&=&X_{GG}^{ph}(q) - X_{FF}(q) \\
\label{eq:XGF}
X_{GF}^{-}&=&X_{GF}(q)- X_{GF}(-q) 
\ee
where the different terms are the following four-dimensional integrals:
\be
\label{eq:Xinit}
X_{GG}^{ph}(q)&=&\frac{1}{i}\int\frac{dk}{(2\pi)^4}G(k)G(k+q)\;\;;\;\;
X_{GG}^{ph}(-q)=X_{GG}^{ph}(q) \\
X_{GG}^{pp}(q)&=&\frac{1}{i}\int\frac{dk}{(2\pi)^4}G(k)G(-k+q) \\
X_{GG}^{pp}(-q)&=&\frac{1}{i}\int\frac{dk}{(2\pi)^4}G(k)G(-k-q) \\
X_{GF}(q)&=&\frac{1}{i}\int\frac{dk}{(2\pi)^4}G(k)F(k+q) \\
X_{GF}(-q)&=&\frac{1}{i}\int\frac{dk}{(2\pi)^4}G(k)F(k-q) \\
X_{FF}(q)&=&\frac{1}{i}\int\frac{dk}{(2\pi)^4}F(k)F(k+q)\;\;;\;\; X_{FF}(-q)=X_{FF}(q)
\label{eq:Xfin}
\ee
\noindent
In the last equations $ G $ and $ F $ are the normal and anomalous proton single particle Greens functions. The quantity  
$ X^{ph}_{n} $ is the analogous of $ X^{ph}_{p} $, but for the non-superfluid neutrons, i.e. in this case only the normal Green' s function appears. Finally $ X^e $ is the relativistic electron Lindhard function \cite{Jan}.\par
The response functions appearing in Eq. (\ref{eq:RPA2}) are labeled according to the operator that appears on the left of the
time-ordered product of Eq. (\ref{eq:Pi}), according to the list (\ref{eq:conf}). 
So the label $ (pp) $ stands for the configuration 
$ (a^\dag(p) a^\dag(p) \,-\, a(p) a(p)) | \, \Psi_0 \, >$, 
while the labels $ (ph,p) $, $ (ph,n) $ and $ (ph,e) $ 
stand respectively for the particle-hole configurations 
$ a^\dag(p) a(p) | \, \Psi_0\, > $,
$ a^\dag(n) a(n) | \, \Psi_0\, > $ and 
$ a^\dag(e) a(e) | \, \Psi_0\, > $.
The same labeling is used for the free response functions $ \Pi_0 $ appearing on the right hand side. For each choice of the configuration on the right of the
the time-order product of Eq. (\ref{eq:Pi}) the free response functions will change properly and the corresponding correlated response functions can be calculated. The subscript $ S $ in the response functions specifies that they are calculated for the spin zero (scalar)  case.  \par 
The analytic expressions for the integrals of Eqs. (\ref{eq:Xinit}-\ref{eq:Xfin}) can be found in ref. \cite{paper3}. 

\end{document}